\documentclass{article}
\newcommand{\bfr}{\begin{flushright}}
\newcommand{\efr}{\end{flushright}}
 
\begin{document}
\title{GAUGE FIELDS ON TORUS AND PARTITION FUNCTION OF STRINGS
}
\author{
Atsushi Nakamula\\
Department of Physics, Tokyo Metropolitan University,\\
Setagaya-ku, Tokyo 158, Japan\\
and\\
Kiyoshi Shiraishi\\
Institute for Nuclear Study, University of Tokyo, \\
Midori-cho, Tanashi,
Tokyo 188, Japan
}
\date{International Journal of Modern Physics {\bf A4} (1989) pp.
389--400
}
\maketitle
\begin{abstract}
In this paper we consider the interrelation between compactified string
theories on torus and gauge fields on it. We start from open string
theories with background gauge fields and derive partition functions by
path integral. Since the effects of background fields and
compactification correlate only through string zero modes, we
investigate these zero modes. From this point of view, we discuss the
Wilson loop mechanism at finite temperature. For the closed string, only
a few comments are mentioned.
\end{abstract}

\section{Introduction}
It is believed that the string theory is one of the promising candidates for the unified
theory.\cite{1} String theories have many different ingredients than
field theories. Firstly, in string theories, there exist infinitely many
heavy particles as string oscillation modes. The existence of these
particles modifies the quantum behavior of the theory. Secondly, since
string theories are often constructed in higher dimensions, the
compactification of extra dimensions is required. As a result, heavy
particles appear in inverse proportion to the size of the internal
space. For they have mass scales about Planck mass, it cannot seem to
produce them directly by the present accelerators.

It can be thought that we will find indirectly, these characteristic phenomena of
string theories only in cosmology; especially in the early universe at high temperature
state, there is a possibility that even the space-time structure is different from now. At
present, it might be able to observe the traces and influences of the early stage of the
universe.

When we consider these possibilities, it is necessary to investigate the relation
between the string theory and the compactification of extra spaces. Further, since string
theories contain many kinds of fields including gauge fields, we have to examine their
own roles.

In this paper we consider the gauge field on torus and its correlation with string
zero-mode components. Then we investigate the (one-loop) partition
function described as path integral in order to understand how to
relate the mass spectra of matter fields with gauge fields.

One of the motivations of this work is to apply the Kaluza-Klein ideas
to string theories. Here, the treatise of background gauge fields and/or
compactified spaces require almost the same prescription as taking
appropriate assumptions on background gauge fields when the
dimensional reduction is done in Kaluza-Klein theory. We will find
``Kaluza-Klein'' excitation mode on each string oscillation mode
spectrum.

Another motivation comes from the interest in the problem of gauge symmetry
breaking in the string theory. String theories formulated in higher dimensions are
generally required to have large gauge symmetries. The idea to derive almost realistic
gauge symmetry when the space is compactified is already used in Kaluza-Klein field
theory. In this mechanism \cite{2} gauge fields as the background field
have an important role; in the string theory, especially its
path-integral approach, it is interesting to know how that mechanism
realizes. In addition, whether ``symmetry breaking'' occurs as ``phase
transition'' or not, and what is the difference from the Higgs
mechanism, are also interesting topics. For the sake of these, it is
necessary to compute the thermo-partition function. In path-integral
formalism, both for field theories and for string theories, the use of
``imaginary-time method'',\cite{3} which is known for
introducing temperature, simplifies the derivation of the
thermo-partition function. Therefore it seems that important results can
be found clearly by considering background gauge fields in string path
integral.

The contents of this paper are as follows. In Sec.~2, we consider the
partition function of open strings with a magnetic flux on
two-dimensional torus. Although the result is not a new one, the
derivation is considerably simplified by the path-integral approach. In
Sec.~3, we investigate the partition function with gauge fields of
nonzero value on torus. This is the case known as Wilson-loop or
Hosotani mechanism,\cite{2} and connected with the spontaneously
symmetry breaking. The dependence of the partition function on
temperature is discussed. In Sec.~4, some comments are given for closed
strings and gauge fields. Sec.~5 is devoted to summary and outlooks.

\section{Magnetic Field and Open String on Two Dimensional Torus}
In this and next sections, we study the open string theory. In this section, a
$U(1)$ gauge field is considered, while non-Abelian gauge field is
studied in the next section.

When we take into account the background electromagnetic field on bosonic open
string theory, the action of two-dimensional world sheet is given as 
follows:\cite{4,5}
\begin{equation}
S=\frac{T}{2}\int
d^2\sigma\sqrt{g}g^{ab}\partial_aX^M\partial_bX_M+iT\oint
ds\partial_sX^MA_M\,,
\label{2.1}
\end{equation}
where $X$'s are ($26$-dimensional) bosonic space-time coordinates, $A_M$
is a background electromagnetic field. $T$ is the string tension.

In this section, we consider the partition function of open strings with a constant
magnetic flux $B$ on two-dimensional torus. Here, it is assumed that the
partition function has already been found in the case with flat
space-time and vanishing magnetic field. Namely, we would like to give
attention to the effect of compactifications and the magnetic field on
the partition function through the string zero-mode.

We restrict ourselves to the case of the ``neutral string'' referred in
Ref.~\cite{4}. Its zero-slope limit yields a Maxwell (electromagnetic)
theory.

We consider the mode expansion of string coordinates, it is known that, by the effect
of the compactifications or the presence of the magnetic field, the eigenfunctions of the
oscillator part are changed only by phase,\cite{4} while their
eigenvalues, which give the mass spectrum, is never changed. So, we give
attention to only zero-mode piece which concerns with ``Kaluza-Klein''
excitation states.

The shape of the world sheet corresponding to the one-loop calculation is a
cylinder-like surface. We assume the path along the $\sigma_2$ coordinate
is closed, and the metric is given as follows:\cite{6}
\begin{equation}
ds^2=d\sigma_1^2+t^2d\sigma_2^2\,.\quad (0\le \sigma_1, \sigma_2\le 1)
\label{2.2}
\end{equation}
($t$ is called the moduli parameter.)

Let $X^1$ and $X^2$ be the torus coordinates, and let $r_1$ and $r_2$
be their radius respectively.
The zero mode parts of them are assumed to satisfy the following condition,
\begin{eqnarray}
X^1(\sigma_1, \sigma_2+1)&=&X^1(\sigma_1, \sigma_2)+2\pi r_1l\,,\nonumber
\\ X^2(\sigma_1, \sigma_2+1)&=&X^2(\sigma_1, \sigma_2)+2\pi r_2 m\, .
\quad(l\mbox{ and } m \mbox{ are integers})
\label{2.3}
\end{eqnarray}

These  conditions indicate  the  fact  that  points  separated  by
$2\pi r_{1,2}$ units in  each direction are identified with each
other, therefore they represent a torus.

Let the following magnetic field exists on the torus,
\begin{equation}
F_{12}=-F_{21}=\nabla_1A_2-\nabla_2A_1\equiv B\, .
\label{2.4}
\end{equation}

Since this field is a kind of ``monopole'' magnetic fields, $B$ is
subject to a quantization condition.\cite{7} However, as seen later, a
conclusion can be derived without that condition.

From the action (\ref{2.1}), it is understood that the following boundary
conditions have to be satisfied,\cite{4}
\begin{equation}
t\partial_{\sigma_1}X^1=iB\partial_{\sigma_2}X^2\,,\quad
t\partial_{\sigma_1}X^2=-iB\partial_{\sigma_2}X^1\,.
\label{2.5}
\end{equation}

From Eqs.~(\ref{2.3}) and (\ref{2.5}), the zero mode components can be
written as follows:
\begin{equation}
\bar{X}^1=\frac{iB}{t}2\pi r_2m\sigma_1+2\pi r_1l\sigma_2\,,\quad
\bar{X}^2=-\frac{iB}{t}2\pi r_1l\sigma_1+2\pi r_2m\sigma_2\,.
\label{2.6}
\end{equation}

Next, let us see the part including gauge fields in the action. 
The line integral of the
gauge field along the boundary can be re-written by the surface 
integral of the field
strength on the world sheet, i.e.,
\begin{equation}
iT\oint
ds\partial_sX^MA_M=-\frac{i}{4}\int
d^2\sigma\varepsilon^{ab}\partial_aX^M\partial_bX^NF_{MN}\,.
\label{2.7}
\end{equation}
Now, we can find the contribution of zero-mode pieces in the partition
function. From (2.1), (2.2), (2.6) and (2 7), we find that the partition
function is proportional to the following factor;
\begin{eqnarray}
& &\sum_{l=-\infty}^\infty\sum_{m=-\infty}^\infty
e^{-\bar{S}}\nonumber \\
&=&\sum_{l}\sum_{m}\exp\left[-T\left\{\frac{1}{2}\int d^2\sigma
\left(t\partial_{\sigma_1}\bar{X}^M\partial_{\sigma_1}\bar{X}_M+
\frac{1}{t}\partial_{\sigma_2}\bar{X}^M\partial_{\sigma_2}\bar{X}_M
\right)\right.\right.\nonumber 
\\
&
&\left.\left.-\frac{i}{4}\int
d^2\sigma\varepsilon^{ab}\partial_a\bar{X}^M
\partial_b\bar{X}^NF_{MN} \right\}\right]\nonumber \\
&=&\sum\sum\exp\left[-T\left\{\frac{1}{2}t\frac{(-B^2)}{t^2}
(2\pi r_2m)^2+\frac{1}{2t}(2\pi r_1l)^2\right.\right.\nonumber \\
& &\left.\left.+\frac{1}{2}t\frac{(-B^2)}{t^2}(2\pi r_1l)^2+\frac{1}{2t}
(2\pi r_2m)^2+\frac{B^2}{2t}[(2\pi r_1l)^2+(2\pi
r_2m)^2]\right\}\right]\nonumber
\\ &=&\sum\sum\exp\left[-\frac{T}{2t}\{(2\pi r_1l)^2+(2\pi r_2m)^2\}
\right]\,.
\label{2.8}
\end{eqnarray}

Then, it turns out to be independent of the magnetic field $B$. The
partition function including the contribution from oscillators is then
given by
\begin{equation}
Z\propto\int_0^\infty\frac{dt}{t}t^{-13}e^{\pi
t}\left[\prod_{n=1}^\infty(1-e^{-\pi nt})\right]^{-24}\sum_{l,m}
\exp\left[-\frac{T}{2t}\{(2\pi r_1l)^2+(2\pi r_2m)^2\}\right]\,,
\label{2.9}
\end{equation}
where $t$ plays a role of the Schwinger parameter of the heat-kernel
method.\cite{8} By using Jacobi's imaginary transformation,\cite{9} we
can find the particle mass spectrum; namely, the partition function is
represented by the integration over $t$ (the heat-kernel integral), and
can be written as the following form,
\begin{equation}
Z\sim\sum_i\int_0^\infty\frac{dt}{t}t^{-D/2}\exp(-tM^2_i)\,,
~(D \mbox{ is the dimension of the spacetime})
\label{2.10}
\end{equation}
where, $M$'s stand for masses of the particles, $\Sigma$ symbolically
represents the summation over the particle species including the
degeneracy. As a result, using Jacobi's imaginary transformation, we get
\begin{equation}
\sum_l\exp\left[-\frac{T}{2t}(2\pi
rl)^2\right]=\sqrt{\frac{t}{2\pi Tr^2}}\sum_l\exp\left[-\pi
t\frac{l^2}{2\pi Tr^2}\right]\,.
\label{2.11}
\end{equation}
Thus it can be read from the form of the partition function that the
mass spectrum is as follows;
\begin{equation}
M^2=2\pi TN'+\frac{l^2}{r_1^2}+\frac{m^2}{r_2^2}\,,
\quad (l\mbox{ and }m\mbox{ are integers})
\label{2.12}
\end{equation}
where $N'$ is the occupation number of oscillators.

Being independent of whether the magnetic field exists or not, the mass spectrum
has a structure that each string-oscillation mode contains Kaluza-Klein excitation
modes. Because we study the ``neutral string'' called in Ref.~\cite{4} in
this section, the fact that the spectrum is independent of magnetic
field is physically rather trivial. It is interesting, however, that the
contribution of string zero-mode under the particular boundary condition
and that from the part of the action including gauge field are cancelled
by each other in the partition function, it is independent of the
quantization condition of the magnetic flux on the compactified space.

In the next section, we will consider the case that the field strength is zero while there
exist nontrivial gauge fields on torus.

\section{Open String Theory and Hosotani Mechanism}
As mentioned in Sec.~1, the method of breaking the gauge symmetries is
an important problem which arises commonly in any unified theories. In
this section, we study the change of particle spectrum by Wilson loop
mechanism (Hosotani mechanism
\cite{2}) with the torus compactification in open string theories. In
addition, we introduce temperature and discuss the possibility of the
phase transition with temperature.

In this section, we consider open superstring theories which contains
$SO(N)$ gauge symmetry. Supersymmetric string action can be expressed by
various forms. Here, we take the covariant one, for example, as in
Ref.~\cite{10} (of open string version). Since we are now concerned with
the part directly connecting gauge fields and zero modes of bosonic
coordinate, we do not write down the total action. The difference
between the part of the action including gauge fields and the previous
one (\ref{2.1}) is that the gauge field is represented as $N\times N$
matrix in the present case. When one calculates the partition function,
the following factor appears:\cite{11}
\begin{equation}
\prod_{boundaries}{\rm Tr } \exp\left[i\oint ds
\partial_sX^M{\bf A}_M
\right]\, .
\label{3.1}
\end{equation}

This form exactly represents the sum of the (gauge-invariant) Wilson-loop elements.
In case of $SO(N)$ symmetry action, string one-loop diagrams have not
only a cylinder but also a twisted band---M\"obius band. Now, we restrict
ourselves to consider the most simple assumption on the gauge field.
Namely, the gauge field for the direction $X^I$ of the torus coordinate
is given by
\begin{equation}
{\bf A}_I=A\left[
\begin{array}{ccc}
0  & -i & \\
i & 0 & \\
 & & 
\end{array}
\right]\,.
\label{3.2}
\end{equation}

Here, in general, it is expected that the symmetry breaking like
$SO(N)\rightarrow SO(N-2)\times U(1)$ occurs. The zero-mode part of the
coordinate can be written by, as the previous section,
\begin{equation}
\bar{X}^I=2\pi rl\sigma_2\, .\quad (l\mbox{ is an integer.})
\label{3.3}
\end{equation}

Substituting (\ref{3.2}) and (\ref{3.3}) into (\ref{3.1}), one can get
the following factor for each diagram.\cite{11}
\begin{eqnarray}
& &\mbox{For the cylinder}: (N+2\{\cos(2\pi l\phi)-1\})^2\,,\nonumber \\
& &\mbox{for the M\"obius strip}: N+2\{\cos(4\pi l\phi)- 1\}\,,
\label{3.4}
\end{eqnarray}
where $\phi=rTA$.

Here, in the trivial case, i.e.~with $\phi=0$, we ought to note that the
factors for a cylinder and M\"obius strip become $N^2$ and $N$
respectively. The partition function is expressed as the summation of
contributions from the cylindrical and M\"obius band diagram, and these
factors agree with the results derived from Chan-Paton factors.\cite{12}

Now, the partition function is, even with the compactification, identically zero
because of supersymmetry. So, here we consider the finite temperature case for an
interesting example. The introduction of temperature has already been achieved by
many people.\cite{10,13} The method can be seen as almost the same
as the ``imaginary time method''\cite{3} in the field theory. By making
the zero-mode part in the time direction be periodic, we introduce the
temperature into the path integral for strings. Namely,
\begin{equation}
\bar{X}^0=\beta n\sigma_2\, ,\quad (n\mbox{ is an integer})
\label{3.5}
\end{equation}
where $\beta$ means the inverse of temperature.

For fermions, we may use the formalism developed in Ref.~\cite{10}. For
the boundary conditions of fermionic coordinates, we take only the
following one corresponding to $n$ in (\ref{3.5}):
\begin{equation}
\theta(\sigma_1, \sigma_2+1)=(-1)^n \theta(\sigma_1, \sigma_2)\,.
\label{3.6}
\end{equation}

This yields just the imaginary time method in superstring version, that is,
``fermion fields change their sign when they are pushed by a period for
the closed time direction''.
The derivation of the partition function can be carried over in
parallel with Ref.~\cite{11}. Let us refer to it for the calculations.

As a final procedure, we sum up two contributions from one-loop diagrams, taking
into account their relative sign. The sign can be determind from the requirement of
existence of $SO(N)$ gauge fields at zero-mass level, for satisfying
unitarity. Then, we obtain the following result.
\begin{eqnarray}
&F&\propto -16T^5\int_0^\infty\frac{dt}{t}(2\pi t)^{-5}\sum_{n=0}^\infty
\exp\left[-\frac{T}{2}\frac{\beta^2}{t}(2n+1)^2\right]\nonumber \\
&\times&
\sum_{l=-\infty}^\infty\exp\left[-\frac{T}{2}\frac{(2\pi
r)^2}{t}l^2\right]\cdot\frac{1}{2}\left[\{N+2(\cos(2\pi\phi l)-1)\}^2
\prod_{p=1}^\infty\left(\frac{1+e^{-\pi
pt}}{1-e^{-\pi pt}}\right)^8\right.\nonumber
\\ & &\qquad\qquad\qquad\left.-\{N+2(\cos(4\pi\phi l)-1)\}
\prod_{p=1}^\infty\left(\frac{1+(-e^{-\pi
t})^p}{1-(-e^{-\pi t})^p}\right)^8\right]\,.
\label{3.7}
\end{eqnarray}

From this partition function, we can find the mass spectrum of the particles by using
Jacobi's transformation as seen in Sec.~2. This is a little more
complicated than the previous one;
\begin{equation}
\sum_{l=-\infty}^\infty\exp\left[-\frac{(2\pi
r)^2Tl^2}{2t}\right]\cos(2\pi\phi l)=\sqrt{\frac{t}{2\pi
Tr^2}}\sum_{l=-\infty}^\infty\exp\left[-\pi t\frac{(l-\phi)^2}{2\pi
Tr^2}\right]\,.
\label{3.8}
\end{equation}

This indicates that Kaluza-Klein excitation modes with the difference by the gauge
field contribution.\cite{14} In the field theory, when there exist gauge
fields on torus, the free energy can be written formally as follows;
\begin{equation}
F\propto\frac{1}{2} {\rm Tr }\ln[D_M^2+M^2]\, .
\label{3.9}
\end{equation}

Here, since covariant derivatives contain gauge fields, we may understand that the
discrete excitation levels on torus originated from these derivatives are shifted by the
contribution from vacuum gauge fields. It is interesting that this local
(``particle-like'') viewpoint connects with a global (``string-like'')
one by the transformation (\ref{3.8}). By the way, we note here that the
concrete expression of the free energy argees exactly with (\ref{3.7})
(see Ref.~\cite{13} etc.). In fact, applying Eq.~(\ref{3.9}) into
(\ref{3.8}), we see the $\phi$ dependence of mass spectrum; 
\begin{eqnarray}
& &\mbox{for even }N':\nonumber  \\
& &M^2=\left\{
\begin{array}{ll}
2\pi TN'+l^2/r^2 & \mbox{with degeneracy }(N-2)(N-3)/2+1 \\
2\pi TN'+(l-\phi)^2/r^2 & \mbox{with degeneracy }2(N-2)\,,
\end{array}
\right.\nonumber \\
& &\mbox{for odd }N':\nonumber \\
& &M^2=\left\{
\begin{array}{ll}
2\pi TN'+l^2/r^2 & \mbox{with degeneracy }(N-2)(N-1)/2+1 \\
2\pi TN'+(l-\phi)^2/r^2 & \mbox{with degeneracy }2(N-2) \\
2\pi TN'+(l-2\phi)^2/r^2 & \mbox{with degeneracy }2\,.
\end{array}
\right.
\label{3.10}
\end{eqnarray}

As above, the patterns of the particle mass spectra modified by the
expectation value of the gauge fields exactly indicate the adjoint or
symmetric representations of $SO(N)$, corresponding to even or odd number
of string oscillators respectively. From other perspective, in
particle-like viewpoint, it corresponds to the fact that in
Eq.~(\ref{3.9}), the covariant derivatives are as follows: 
\begin{eqnarray}
& &\mbox{If }\lambda_A \mbox{ is in the adjoint representation}:
\nonumber \\
& &\qquad\qquad\qquad\qquad\qquad\qquad
D_M\lambda_A\sim\partial_M\lambda_A+i[{\bf A}_M,
\lambda_A]\,.\nonumber
\\ & &\mbox{If }\lambda_S \mbox{ is in the symmetric representation}:
\nonumber \\
&
&\qquad\qquad\qquad\qquad\qquad\qquad
D_M\lambda_S\sim\partial_M\lambda_S+i\{{\bf A}_M,
\lambda_S\}\,.
\label{3.11}
\end{eqnarray}

It is interesting that the factor given by the sum of two different world sheet diagrams,
cylinder and M\"obius band, turns out to give correctly the symmetric
patterns of the group.

The ``free energy'' (\ref{3.7}) may be regarded as a potential of $\phi$.
Now we proceed in this line.

We find that $\phi=0$ (modulo $1$) is the minimum of the free energy even
at nonzero temperature (of course, at zero temperature it closes to be
identically zero) in this model.
Thus we cannot have naive image of phase transitions from this model. Unfortunately,
the same may be true whenever we consider the free energy with the Wilson-loop on
torus. Before replacing the model with more complicated ones, let us investigate its
temperature dependence in order to get a physical insight.

First of all, we note that when temperature $(\beta^{-1})$ is extremely
smaller than the square root of the string tension $(T^{1/2})$, the
behavior of the free energy agrees with that obtained by usual field
theoretical techniques because we can neglect the excitation of string
oscillations. Similar to Ref.~\cite{14}, the potential may be regarded as
a summation of the form $\cos(2\pi l\phi)$, so we can only think about
its coefficient for each $l$. In short, it is important that the cosine
functions do not contain the parameter $t$ of the integration.

To proceed in detail, in the model with supersymmetry as considered here, the
analysis of the potential at low temperature is difficult because of the canceHation of
the coefficients which assume large contributions in the potential. Without a numerical
calculation, we investigate it by the following method. Firstly, the extremum of the
potential can be at $\phi=0$ or $1/2$, because the potential is given by
the sum of trigonometric functions. Next, we should consider the
difference of the potential at $\phi=0$ and $\phi=1/2$. The difference is
reduced to the following summation,
\begin{equation}
-\sum_{l:odd}\sum_{n:odd}\frac{1}{[(2\pi rl)^2+(\beta n)^2]^5}\,.
\label{3.12}
\end{equation}

As far as the string excitation modes can be neglected, the difference of the potentials
is proportional to this summation. Trivially, (\ref{3.12}) is negative,
so it is expected that at low temperature $(\beta^{-1}\ll T^{1/2})$ the
minimum of the free energy is realized when $\phi=0$.
Especially in the case of $\beta^{-1}\ll r^{-1}$, the difference of the
free energy at $\phi=0$ and $\phi=1/2$ behaves like $\beta^{-9}$.

Next, we consider the high temperature case. In the string theory, there exists the
critical temperature \cite{15} at which several physical quantities
diverge. But the difference of the potential as studied above remains
finite. At the critical temperature, the difference of the energy at
$\phi=0$ and $\phi=1/2$ can be evaluated as follows,
\begin{eqnarray}
\Delta F&\equiv& F(\phi=0)-F(\phi=1/2)\nonumber \\
&\sim& -(N-2)(2\pi)^{-5}T^4\sum_{l=0}^\infty\sum_{n=0}^\infty
\frac{1}{(2\pi r)^2(2l+1)^2+4\beta_c^2n(n+1)}\,,
\label{3.13}
\end{eqnarray}
where $\beta_c=(4\pi/T)^{1/2}$ is the inverse of the critical
temperature.

Therefore, we can find that even if string oscillations come into the calculation,
the locations of the minima of the potential (free energy) are never shifted at high
temperature.

From the brief investigation given here, we can guess that the minimum of
the potential is independent of the temperature in general models.
Further, even at high temperature, since the critical temperature is the
order of the lowest string excitation mode, we will understand that at
``medium'' temperature the ``correction'' by string oscillation gives
little influence. 

It may be conceivable to see other mechanism that is
more complicated but possible to break gauge symmetry for instance
compactifications on orbifolds etc.; to study them will be a future
problem.

\section{Comments on Closed Strings}
The partition function for closed strings on torus is definitely given by Sakai and
Senda.\cite{16} They have used Hamiltonian (canonical) formalism;
however the same result can be obtained by the path integral method. In
this method, we impose a Kaluza-Klein-like assumption on the space-time
metric $G_{MN}$; then, we can obtain half of the zero-mode gauge fields,
corresponding to the Cartan subgroup of the gauge group generated by
compactification, while the rest is included by consideration of the
antisymmetric tensor $B_{MN}$. It ls symbolically written as follows.

Gauge fields of zero modes $A$ and $B$ corresponding to the Cartan
subgroup $H\times H'$ of the gauge field $G\times G'$ are given by
\begin{equation}
A_\mu\sim G_{\mu I}\quad\mbox{and}\quad B_\mu\sim B_{\mu I}\,,
\label{4.1}
\end{equation}
where $I$'s are the indices of torus-compactified dimensions.

These representations can be used to introduce chemical
potential.\cite{17}

Now, we may consider Wilson-loops in terms of these gauge fields. As a concrete
example, we will investigate bosonic strings in twenty-six dimensions and the case that
one dimension is compactified on torus. When the size of the torus takes a special value,
we can regard it as the theory possessing $SU(2)\times SU(2)$ gauge
symmetry in twenty-five dimensions.\cite{16} And then letting another
dimension be torus, we can set $A$ and $B$ have nonzero values on it.
This formulation leads to the nontrivial Wilson-loop elements.

Regarding it as an original twenty-six dimensional theory, the existence of nonzero
$A$ and $B$ is equivalent to the possibility of two-dimensional torus to
take various forms and nonzero values of antisymmetric tensor fields on
it. As shown by Sakai and Senda, with the metric and antisymmetric
tensor fields of special values, the gauge symmetry can be extended to
$SU(3)\times SU(3)$.

Considering in twenty-five dimension again, it can be said, ``the gauge
symmetry is extended' by the Wilson-loop mechanism.\cite{11} Of course,
even by taking into account the Kaluza-Klein excitation of gauge fields,
it never occurs in usual field theories. 

It was not shown explicitly by
Sakai and Senda, however, that the existence of antisymmetric tensor
fields is a necessary condition for what is shown above to occur.
However a further essential point will be that, in the closed string
theory, quantum numbers are restricted by the freedom in
reparametrization of string coordinates. The constraint can be, although
not written here, derived from path integral formalism of partition
function. (See Ref.~\cite{13}.) We should note that the constraint on
quantum numbers is never changed with any values of $A$,
$B$, and the radius of the torus. We must give attention to the fact that
when we introduce temperature by the imaginary time method, the time
axis can be regarded as torus while there is no restriction by coordinate
reparametrization in its direction.

This restriction and the existence of tachyon are necessary conditions
for the mechanism of gauge-symmetry extension in the closed string
theory. In the heterotic string theory, since gauge fields can be seen
as obtained from compactification, it must be careful when we consider
Wilson-loop (Hosotani) mechanism. The practical calculation of the
free energy with constant gauge fields is straightforward in the path
integral formalism, because many authors have considered such background
fields in the context of the study in the string theory with various
gauge symmetries.\cite{18}

Consequently, when we are concerned with the Wilson-loop or vacuum gauge fields,
in case of the open string the particle viewpoint remains true, since the interaction with
external gauge fields is restricted at the edges of world sheet; meanwhile in the closed
string theory there appears essentially characteristic property of the string theory.
However, this is the case in one loop level. Even in the open string
theory the `stringy' nature will appear in higher loop levels; the
corresponding diagrams might be the form containing closed strings in
those levels. 

In the closed string theory, we must take care of the
winding soliton states. Thus the evaluation of the free energy becomes
difficult when the scale of compactification is small. We suppose the
treatment of the Wilson-loop mechanism can be easy in the ``canonical''
way that, roughly speaking, we construct explicitly mass and charge
eigenstates. The line of the thought will be developed in separate
publications.

\section{Summary}
In this paper, we studied how string partition function is affected by the part
containing gauge fields on torus and string zero mode in the path integral formalism.
Especially, because the size of the torus and the temperature are
represented by zero mode of strings, the partition function can be given
through a simple derivation.

In future, we would like to consider
whether the ``Kaluza-Klein'' idea is applicable to even the case fully
affected by characteristic properties in string theory, such as, the
closed string theory on orbifolds and Wilson-loops on it \cite{19} and
the superstring theory in lower dimensions than ten.\cite{20}

As for the dependence of the partition function on temperature, it will be interesting
to consider compactification of nonsupersymmetric string
theories.\cite{21} In these kind of theories it is known, for example,
that the partition function at zero temperature, i.e. the cosmological
constant, asymptotically goes to zero with the radius of compactified
torus.\cite{22} Such models will be expected to lead a new interest
about cosmological evolution of the space-time structure and (gauge)
symmetries. These models are described by the closed string theories
and have some complexities; nevertheless, when we consider the early
universe, it should be required to search for the partition function at
finite temperature and to investigate its behavior.

Finally, it must be mentioned the development of four-dimensional string
theories.\cite{23} However, the theories with extra dimensions should be
investigated with sufficient interest, because of
(1) the possibility to solve the problems in (four-dimensional)
cosmology,\cite{24} (2) the requests from quantum cosmological
models,\cite{25} (3) to supply a mechanism for supersymmetry
breaking,\cite{26} and so on; further they might have many more
important roles.

\section*{Acknowledgments}
We thank T. Hori for reading this manuscript. One of us (K.S.) thanks
Iwanami F\=ujukai for financial support.


\end{document}